\documentclass[preprint,aps,12pt,preprintnumbers,nofootinbib]{revtex4-2}
\usepackage{xcolor}
\usepackage{graphicx}
\usepackage{epstopdf} 
\usepackage{braket}
\usepackage{amssymb,amsmath,amsfonts,comment,latexsym,graphicx,epstopdf,float}
\usepackage{color}
\usepackage{fancyvrb}
\usepackage{chngcntr}
\usepackage{etoolbox}
\usepackage{ulem}
\usepackage{array}
\usepackage{slashed}
\usepackage{multirow}

\usepackage[colorlinks=true,citecolor=blue,linkcolor=red,filecolor=cyan,urlcolor=magenta]{hyperref}

\usepackage{float}

\newcommand{\be}{\begin{equation}}
\newcommand{\ee}{\end{equation}}
\newcommand{\ba}{\begin{eqnarray}}
\newcommand{\ea}{\end{eqnarray}}

\newcommand{\grts}{\raise.3ex\hbox{$>$\kern-.75em\lower1ex\hbox{$\sim$}}}
\newcommand{\lets}{\raise.3ex\hbox{$<$\kern-.75em\lower1ex\hbox{$\sim$}}}

\newcommand{\dd}{\text{d}}

{\catcode`\|=\active\gdef\Braket#1{\left<\mathcode`\|"8000\let|\bravert 
{#1}\right>}}

\def\bravert{\egroup\,\vrule\,\bgroup}

\usepackage{color}
\usepackage{amssymb,amsmath,amsfonts}
\usepackage{epstopdf} 
\usepackage{braket}

\usepackage{autobreak}

\begin{document}
%
%
\title{\vspace*{0.5in} 
Asymptotically safe dark matter with gauged baryon number 
\vskip 0.1in}
\author{Jens Boos}\email[]{jboos@wm.edu}
\author{Christopher D. Carone}\email[]{cdcaro@wm.edu}
\author{Noah L. Donald}\email[]{nldonald@wm.edu}
\author{Mikkie R. Musser}\email[]{mrmusser@wm.edu}
\affiliation{High Energy Theory Group, Department of Physics,
William \& Mary, Williamsburg, VA 23187-8795, USA} 
%
%
\date{October 4, 2022}
\begin{abstract}
We consider the inclusion of TeV-scale, fermionic dark matter in an asymptotically safe model of gauged baryon number that has been recently proposed
[Phys.\ Rev.\ D \textbf{106}, no.~3, 035015  (2022)].  The new gauge boson serves as a portal between the dark and the visible sectors.  The 
range of the baryon number gauge coupling and the kinetic mixing between baryon number and hypercharge 
are constrained by the requirement that nontrivial ultraviolet fixed points are reached.  We show that this asymptotically safe dark matter model can achieve 
the correct dark matter relic density while remaining consistent with direct detection bounds.
\end{abstract}
\pacs{}

\maketitle
\newpage

\section{Introduction} \label{sec:intro}

Quantum field theories that are free of Landau poles are interesting since they allow perturbative extrapolation to arbitrarily high energy scales.  
Couplings in such theories can flow either to vanishing or non-vanishing ultraviolet (UV) fixed points.  If at least some of the fixed 
points are non-vanishing, the theory is said to be asymptotically safe.  Asymptotic safety was proposed by Weinberg~\cite{weinberg00} to 
explain how a quantum field theory of gravity could be predictive, even if it is perturbatively nonrenormalizable; see also 
Refs.~\cite{Berges:2000ew,Niedermaier:2006wt}. Nonrenormalizable theories have an infinite-dimensional parameter space, with a 
coupling for every divergent amplitude in the theory. The requirement of asymptotic safety, however, can reduce this to a 
finite-dimensional subspace, called the ultraviolet critical surface. Within the paradigm of asymptotic safety, it is postulated that physically sensible theories can only be defined in this 
subspace.   Asymptotically safe nonrenormalizable theories can therefore be predictive and serve as viable alternatives to 
renormalizable ones.  For a review, see Ref.~\cite{Percacci:2011fr}.

The requirement of asymptotic safety can also reduce the dimensionality of the parameter space of a renormalizable theory.  This 
can be of value in formulating extensions of the standard model, which typically involve a plethora of new particles and otherwise 
underconstrained couplings.  Asymptotic safety has been used as a principle to restrict the parameter space of a number of 
beyond-the-standard-model scenarios~\cite{Boos:2022jvc,Reichert:2019car,Hiller:2019mou,Kowalska:2020zve,Hamada:2020vnf,Domenech:2020yjf,Grabowski:2018fjj,Kowalska:2022ypk,
Kowalska:2020gie,Bause:2021prv,Chikkaballi:2022urc,Kwapisz:2019wrl,Hiller:2020fbu,Wang:2015sxe,Barducci:2018ysr}, including models of dark 
matter~\cite{Reichert:2019car,Kowalska:2020zve,Hamada:2020vnf}, of new contributions to the muon anomalous magnetic 
moment~\cite{Hiller:2019mou,Kowalska:2020zve}, of modified Higgs, neutrino~\cite{Domenech:2020yjf,Grabowski:2018fjj,Kowalska:2022ypk} and gauge
sectors~\cite{Wang:2015sxe,Kwapisz:2019wrl}, of $B$-meson anomalies~\cite{Kowalska:2020gie,Bause:2021prv,Chikkaballi:2022urc} and of new TeV-scale physics with 
collider physics implications~\cite{Hiller:2020fbu}.  

In the present work, we follow up on a model of asymptotically safe gauged baryon number proposed in Ref.~\cite{Boos:2022jvc}.   There is an 
extensive literature on the possibility of gauged baryon number; for discussion of the motivations and phenomenology, see, for example, 
Refs.~\cite{Carone:1994aa,Carone:1995pu,FileviezPerez:2010gw,Duerr:2013dza,Duerr:2013lka,FileviezPerez:2014lnj,FileviezPerez:2019jju}.
The most important parameters for determining the properties of the new U(1) gauge boson are the gauge coupling $g_B$, the gauge boson 
mass $m_B$, and a coupling $\tilde{g}$ that determines the kinetic mixing between hypercharge and the U(1)$_B$ gauge group.  In the absence of this 
mixing, the tree-level couplings of the new gauge boson to standard model matter fields are entirely leptophobic;  the degree to which 
the model deviates from leptophobia is determined by the kinetic mixing parameter, making its value of critical importance in determining the phenomenology of the model.  
One of the benefits of an asymptotically safe version of the gauged baryon number scenario is that the kinetic mixing is fixed in terms of $g_B$ at the 
TeV scale due to the constraints on the couplings in the deep UV.   This leads to greater predictivity.   Reference \cite{Boos:2022jvc} mapped out the 
ultraviolet fixed points in a simple model of gauged baryon number and discussed the phenomenological consequences, assuming that the scale of 
new physics (including new fermions to cancel anomalies) was around $1$~TeV.     

The model of Ref.~\cite{Boos:2022jvc} did not include a dark matter candidate, an omission that we will remedy here. In addition to the TeV-scale particle content included to cancel anomalies, we add a fermion, $\chi$, that is vector-like under U(1)$_B$ and carries no other gauge quantum numbers:
\begin{equation}
\chi_L \sim \chi_R \sim (1,1,0,1/6) \,\,\, .
\label{eq:chiLR}
\end{equation}
Here, the first two numbers shown are the dimensionalities of the SU(3)$_C$ and SU(2)$_W$ representations, while the last two are the U(1)$_1$ and 
U(1)$_B$ charges.   We work with the grand unified theory (GUT) normalization of hypercharge, {\it i.e.}, $g_Y = \sqrt{3/5} \, g_1$ where $g_Y$ is
the hypercharge coupling of the standard model; $g_B$ is normalized so that the baryon number of a nucleon is $+1$.   The baryon number charge assignment
in Eq.~(\ref{eq:chiLR}) renders the $\chi$ field stable, as we explain in the next section, and allows the U(1)$_B$ gauge field to serve as a portal between the dark and the visible sectors. Other work on 
such ``baryonic'' dark matter candidates appears in Refs.~\cite{Duerr:2013lka,FileviezPerez:2014lnj,FileviezPerez:2019jju}.  Aside from differences in the particle content and charge 
assignments that we assume, our approach differs in that we work in the framework of asymptotic safety where both the allowed
range of $g_B$ and the value of the kinetic mixing
parameter $\tilde{g}$ are constrained by the ultraviolet boundary conditions on the theory.   With $\tilde{g}$ predicted in terms of $g_B$, we include leptonic channels
in the new gauge boson decay width and the dark matter annihilation cross section, without introducing dependence on an additional free parameter.
In addition,  our calculation of the relic density incorporates
a relativistic treatment of the thermally averaged dark matter annihilation cross section times relative velocity, which is expected to be more accurate for the near-resonant annihilation~\cite{Gondolo:1990dk} that we encounter in the present study.

This letter is organized as follows. In Sec.~\ref{sec:two} we discuss the choice in Eq.~(\ref{eq:chiLR}), how dark matter stability is assured, and how the
fixed-point structure of the model of Ref.~\cite{Boos:2022jvc} is affected by the additional particle content. In Sec.~\ref{sec:three}, we study  the dark matter relic density, and in Sec.~\ref{sec:four} we consider the direct detection bounds for points in model parameter space where the correct relic density is 
obtained. We summarize our conclusions in Sec.~\ref{sec:conc}.

\section{Gauged baryon number model} \label{sec:two}

The model of Ref.~\cite{Boos:2022jvc} includes the U(1)$_B$ gauge boson $B{}_\mu$, a scalar field $\phi$ with baryon number $+1$, and a number of new fermions
that are introduced to cancel gauge and gravitational anomalies.  The new gauge boson and fermions become massive when $\phi$ acquires a vacuum 
expectation value (vev); we assume their mass spectrum lies at or above $1$~TeV.  The charge assignments of the fields can be found in Ref.~\cite{Boos:2022jvc}; what is 
important here is that the magnitudes of the baryon number charges $|Q_B|$ are either $0$, $1/3$ or $1$.   Under a U(1)$_B$
phase rotation $\exp(i \,Q_B \, \alpha)$, all of these fields are left invariant in the case where $\alpha= 6 \pi$.   On the other hand, the field $\chi$ in Eq.~(\ref{eq:chiLR}) changes
sign under that action of the same group element.   This establishes that there is a $Z_2$ symmetry which is a subgroup of U(1)$_B$ and that remains unbroken after spontaneous symmetry
breaking.   Since the $\chi$ field is the only field that is odd under this symmetry, its stability is guaranteed, making it a potential dark matter 
candidate. The fact that the stabilizing symmetry is a subgroup of a gauge symmetry renders it safe from violation by any possible quantum gravitational effects.  

The hypercharge and U(1)$_B$ gauge fields can mix through their kinetic terms, 
\begin{equation}
{\cal L} \supset -\frac{1}{4} F^{\mu\nu}_Y F_{\mu\nu}^Y - \frac{1}{4} B^{\mu\nu} B_{\mu\nu} + \frac{\epsilon}{2} B^{\mu\nu} F_{\mu\nu}^Y \,\,\, .
\label{eq:u1kin1}
\end{equation}
We follow a standard approach of working in a basis where the kinetic terms are diagonal and canonically normalized, but where the gauge-covariant derivative for a generic 
field $\Psi$ takes the form~\cite{Boos:2022jvc,Reichert:2019car,Wang:2015sxe}
\begin{equation}
D_\mu \Psi = [\partial_\mu  - i \, g_B B_\mu Q_B -i \, (g_1 A^Y_\mu + \tilde{g} \, B_\mu) Q_1 ]\, \Psi \, .
\label{eq:covd}
\end{equation}
Here, $Q_B$ and $Q_1 = \sqrt{3/5} \, Q_Y$ denote the baryon number and hypercharges of $\Psi$, respectively, and $\tilde{g} = \epsilon \, g_1 / \sqrt{1-\epsilon^2}$.

The one-loop $\beta$ functions for $g_1$, $g_B$ and $\tilde{g}$ (computed using PyR@TE 3~\cite{pyrate}) are
\begin{align}
\begin{split}
\beta^{(1)}(g_1) &= \frac{77}{10}g_1^3 - \hat{f}_g g_1 \, \theta(\mu-M_\text{Pl}) \, , \\
\beta^{(1)}(g_B) &= \frac{298}{27}g_B^3-\frac{16}{3}g_B^2\tilde{g}+\frac{77}{6}g_B\tilde{g}^2-\hat{f}_g g_B \, \theta(\mu-M_\text{Pl}) \, , \\
\beta^{(1)}(\tilde{g}) &= \frac{77}{6}\tilde{g}^3 - \frac{16}{3}\tilde{g}^2 g_B +\frac{298}{27}\tilde{g} g_B^2 +\frac{77}{5}\tilde{g} g_1^2 -\frac{16}{5}g_1^2g_B-\hat{f}_g\tilde{g} \, \theta(\mu-M_\text{Pl}) \, .
\end{split}
\end{align}

\begin{figure}[t]
\centering
\includegraphics[width=0.4\textwidth]{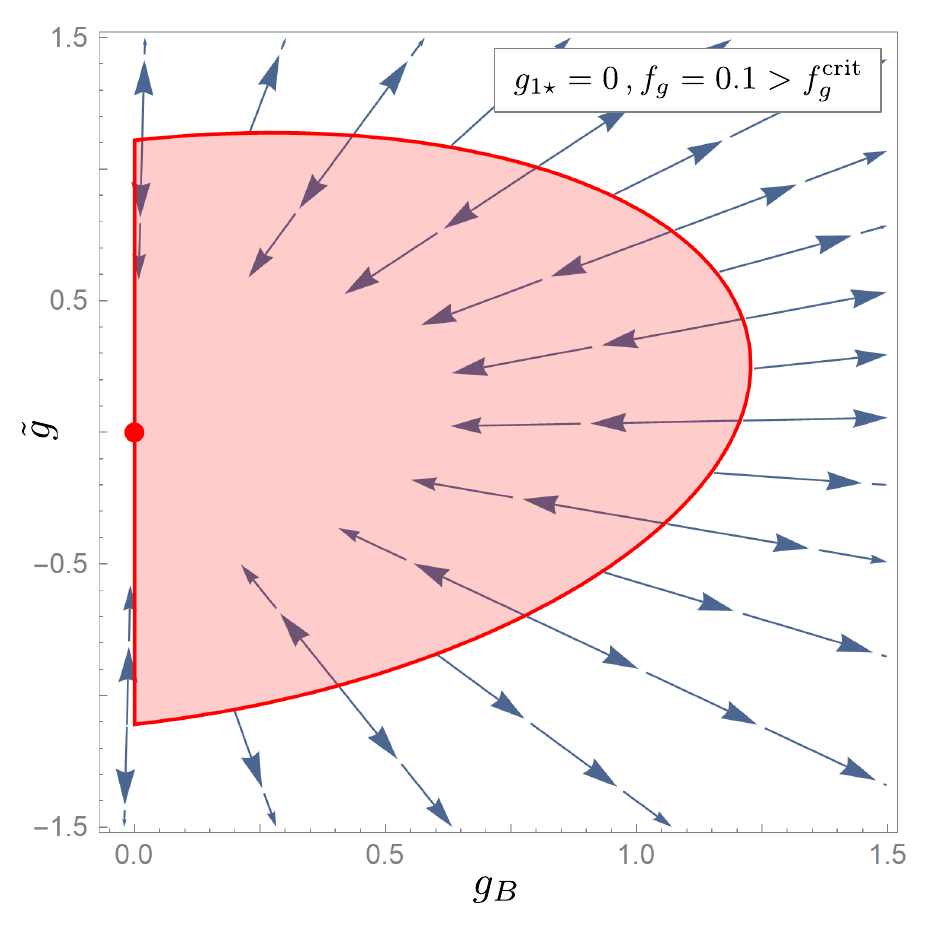}
\includegraphics[width=0.4\textwidth]{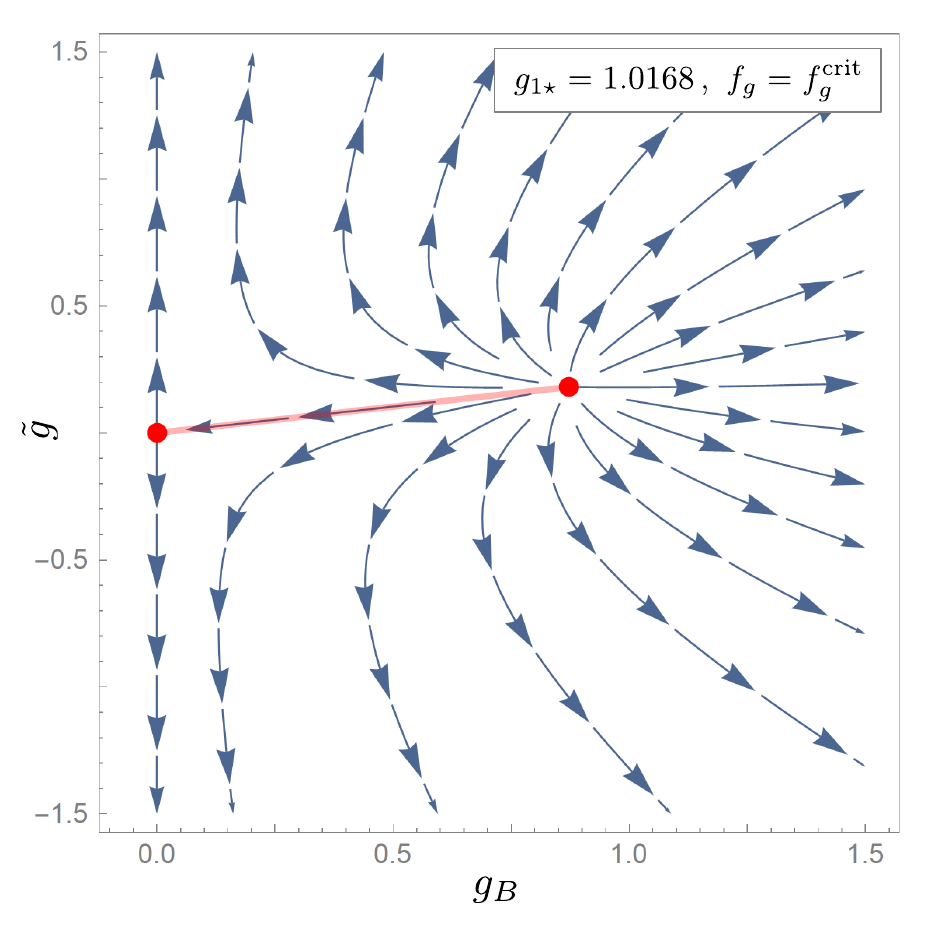}
\caption{Visualization of the RG flow in the $g_B\tilde{g}$-plane. The inside of the ellipse and the interior of the line segment are driven to the Gaussian fixed point $(g_B,\tilde{g}) = (0,0)$; the ellipse's boundary and the line's right endpoint are non-trivial fixed points. In the conventions adopted in this letter, the arrows on the flow lines point towards the UV.}
\label{fig:rgflow} 
\end{figure}

We use the notation $\beta(g) = \beta{}^{(1)}(g)/(4\pi)^2$ and $\hat{f}_g \equiv (4\pi)^2f_g$, for convenience. The inclusion of gravitational correction terms to the gauge coupling $\beta$ functions, with a universal
parameter $\hat{f}_g$, is motivated by functional renormalization group (RG) calculations~\cite{Reichert:2019car,Kowalska:2020zve,Wang:2015sxe}.  
(For an alternative approach towards realizing asymptotic safety, see Ref.~\cite{largeNf}.)  Since the RG running of $g_1$ decouples from that of $g_B$ and $\tilde{g}$, we will distinguish between two different fixed point scenarios, corresponding to the solution of
\begin{equation}
\left( \frac{77}{10}g_{1\star}^2-\hat{f}_g \right)g_{1\star}=0 \, .
\end{equation}
A non-trivial, interacting fixed point is obtained provided that $\hat{f}_g$ has a critical value
$\hat{f}_g^\text{crit}=\frac{77}{10} \, g_{1\star}^2$; in this case, $g_1$ remains constant and nonvanishing above the Planck 
scale, with $\hat{f}_g^\text{crit} \approx 7.9610$ to match the experimental value of $g_1$ at the electroweak scale~\cite{Boos:2022jvc}. For $\hat{f}_g$ larger than the 
critical value, the gravitational term drives $g_1$ to a trivial, Gaussian fixed point.  In either case, the requirement that $g_1$ reaches a 
fixed point constrains the evolution of the remaining couplings $g_B$ and $\tilde{g}$.   Their flow as one evolves to higher renormalization scales is shown graphically in Fig.~\ref{fig:rgflow}.

When $g_1$ flows to its Gaussian fixed point, the locus of fixed points in the $g_B \tilde{g}$-plane is constrained to the ellipse $E(g_{B\star},\tilde{g}_\star, \hat{f}_g)=0$, where
\begin{equation}
E(g_{B},\tilde{g}, \hat{f}_g)=\frac{77}{6}\tilde{g}^2+\frac{298}{27}g_B^2-\frac{16}{3}g_B\tilde{g}-\hat{f}_g \, .
\end{equation}
Points inside this ellipse flow toward $(g_{B\star},\tilde{g}_\star)=(0,0)$, while points outside flow to infinite radius. Loosely speaking, the gravitational correction 
factor $f_g$ sets the size of this ellipse.  On the other hand, when $g_1$ flows to its interacting fixed point, there are two $g_B \tilde{g}$ fixed points 
connected by a line segment in the $g_B \tilde{g}$-plane,
\begin{equation}
\tilde{g}=\frac{16}{77} \, g_B \, .
\end{equation}
The end point $(g_{B\star},\tilde{g}_\star)=(0.87145,0.18108)$ is an unstable fixed point, while points on the interior of the line segment flow to a trivial fixed point at $(g_{B\star},\tilde{g}_\star)=(0,0)$.

The inclusion of the dark matter particle $\chi$ changes the $\beta$ functions from those given in Ref.~\cite{Boos:2022jvc}.  However,  the numerical effects are
small and the pattern of fixed points and flow lines remains qualitatively unchanged.  Given the multitude of choices for ultraviolet fixed points, we will limit our consideration to what are
plausibly three representative cases:
\begin{itemize}
\item[(ia)] Interacting $g_1$ fixed point:  One end of the line segment shown in Fig.~\ref{fig:rgflow} is an unstable fixed point
with  $(g_{1\star},g_{B\star},\tilde{g}_\star)=(1.0168,0.87145,0.18108)$ and $\hat{f}_g^\text{crit} \approx 7.9610$. The requirement of reaching 
an unstable fixed point leads to the greatest predictivity in the low-energy theory.    At 1 TeV, the couplings are 
$(g_1,g_B,\tilde{g}) = (0.46738, 0.40049, 0.083219)$.  Note that $\tilde{g} = 16/77 \, g_B$ is preserved by the RG flow.
\item[(ib)] Interacting $g_1$ fixed point:  Choosing a point on the interior of the line segment with $\tilde{g} = 16/77 \, g_B$ at 1 TeV
and $\hat{f}_g^\text{crit} \approx 7.9610$  again yields an interacting fixed point for $g_1$, but $g_B$ and $\tilde{g}$ now flow to Gaussian fixed points.
For this example, we take $g_B=0.2$, {\it i.e.}, $(g_1,g_B,\tilde{g}) = (0.46738,0.20000, 0.041558)$ at 1~TeV; this flows to
$(g_{1\star},g_{B\star},\tilde{g}_\star)=(1.0168,0,0)$. 
\item[(ii)] Gaussian $g_1$ fixed point: We choose $f_g =0.1$, below the critical value, so that $g_1$ flows to a Gaussian fixed point.   Choosing 
a point on the ellipse provides nontrivial fixed points for $g_B$ and $\tilde{g}$.  For easy comparison to case (ib), we choose a solution for
which $g_B=0.2$ at 1 TeV:  we assume $(g_{1\star},g_{B\star},\tilde{g}_\star)=(0,0.209651, 1.13654)$ which leads to the 1~TeV values 
$(g_1,g_B,\tilde{g}) = (0.46738,0.20000,0.15067)$.
\end{itemize}
In each of these cases, at least one coupling flows to a non-trivial fixed point, corresponding to an asymptotically safe scenario. In what follows, we will evaluate the constraints
on the model parameter space from the dark matter relic density and direct detection bounds.

\section{Relic density} \label{sec:three}
To good approximation, the dark matter $\chi$ remains in thermal equilibrium as long as the annihilation rate to standard model particles 
exceeds the expansion rate of the universe. Dark matter annihilation to standard model fermions via exchange of the U(1)$_B$ gauge boson provides the dominant contribution to the annihilation cross section. Given our assumption  that the new fermions in the model have TeV-scale masses, we neglect the mass of standard model fermions, aside from that of the top quark. The cross section for annihilation into a standard model fermion $f$ is given by 
\begin{align}
\sigma\left(\chi\bar{\chi} \rightarrow f\bar{f} \right) = \frac{N_c \, g_B^4}{1728 \, \pi} \frac{1}{s} \sqrt{\frac{s-4 \, m_f^2}{s-4 \, m_\chi^2}}\left(s + 2\, m_\chi^2 \right) \left[ \frac{C_V^2(s + 2 \,m_f^2) + C_A^2(s-4 \, m_f^2) }{ (s-m_B^2)^2+\Gamma^2 \, m_B^2} \right]\, .
\end{align}
Here, $N_c$ is the number of colors, $m_f$, $m_\chi$, and $m_B$ are the masses of the standard model fermion, the dark matter particle, and the U(1)$_B$ gauge boson, respectively, 
and $\Gamma$ is the gauge boson decay width. The partial decay width to an $f \overline{f}$ final state is given by 
\begin{align}
\Delta \Gamma\left( B \rightarrow f\bar{f} \right) = \frac{N_c \, g_B^2 \, m_B}{48 \, \pi}  \sqrt{1-\frac{4m_f^2}{m_B^2}}\left[ C_V^2 \left(1 + \frac{2 m_f^2}{m_B^2}\right) + C_A^2 \left(1 - \frac{4 m_f^2}{m_B^2}\right) \right] \, .
\end{align} 
The coefficients $C_V$ and $C_A$ are vector and axial-vector couplings in units of $g_B$. One finds numerically that $|C_V|$ is given by $0.8008$, $0.6398$, $0.2414$, and $0.0805$ for up-type quarks, down-type quarks, charged leptons, and neutrinos, respectively, whereas the $|C_A|$ are all $0.0805$.

Dark matter falls out of thermal equilibrium at the freeze-out temperature $T_f$, which we determine by the condition
\begin{equation}
\frac{\Gamma}{H(T_f)}\equiv \frac{n_\chi^{EQ}\langle\sigma v\rangle}{H(T_f)} \approx 1 .
\end{equation}
Here, $n_\chi^\text{EQ}$ is the equilibrium number density and $H(T)=1.66 \sqrt{g_{*}} \,T^2/M_\text{Pl}$ is the Hubble parameter for a radiation dominated universe written in terms of the number of relativistic degrees of freedom, $g_{*}$, and the Planck mass $M_\text{Pl} = 1.22 \times 10^{19}$ GeV.  For a radiation dominated universe, it is appropriate to assume the non-relativistic equilibrium number density
\begin{equation}
n_\chi^\text{EQ}=2\left(\frac{m_\chi T}{2 \pi}\right)^{3/2}e^{-m_\chi/T}.
\end{equation}
A relativistic treatment of the thermally averaged annihilation cross section times relative velocity is given by~\cite{Gondolo:1990dk}
\begin{equation}
\langle \sigma v \rangle = \frac{1}{8 m_\chi^4 T K_2^2\left(\frac{m_\chi}{T}\right)}\int_{4 m_\chi^2}^{\infty} \dd s \, \sigma_{tot} \times (s-4m_\chi^2)\sqrt{s}K_1\left(\frac{\sqrt{s}}{T}\right) \, ,
\end{equation}
where the $K_i$ are modified Bessel functions of order $i$.  For the sufficiently large freeze-out temperatures considered in this analysis, the ratio of the equilibrium number density to the entropy density at freeze-out, $Y_f$, is given by
\begin{equation}
Y_f = 0.145 \frac{g}{g_{*}}x_f^{3/2}e^{-x_f} \, ,
\end{equation}
where $x_f \equiv m_\chi/T_f$ and $g=4$ is the number of internal degrees of freedom of the dark matter particles plus antiparticles.  This ratio at freeze-out can then be propagated to the ratio $Y_0$ at the present temperature of the universe, 
\begin{equation} 
\frac{1}{Y_0}=\frac{1}{Y_f} + \sqrt{\frac{\pi}{45}}M_\text{Pl}m_\chi \int_{x_f}^{x_0} \dd x \frac{\sqrt{g_{*}}}{x^2} \frac{\langle \sigma v \rangle}{2} \, ,
\end{equation}
where the factor of $1/2$ takes into account that annihilation only occurs between a dark matter particle and its antiparticle, while the $Y_i$ in this expression include both~\cite{Gondolo:1990dk}.

The dark matter relic density is then given by
\begin{equation}
\Omega_\text{D} h^2 \approx \frac{2.8 \times 10^8}{\text{GeV}} Y_0 \, m_\chi \, .
\end{equation}
We now compute the relic density in each of the previously described cases (ia), (ib), and (ii). This analysis relies on numerical integration which can be performed to high accuracy. Since the couplings are fixed by the asymptotic safety criterion, the only free 
parameters entering this analysis are the dark matter mass $m_\chi$ as well as the U(1)$_B$ gauge boson mass $m_B$. The latter is assumed to be in the TeV-range, comparable to the masses of the other heavy fermions; see Ref.~\cite{Boos:2022jvc} for details.

\begin{figure}[!h]
\centering
\includegraphics[width=0.6\textwidth]{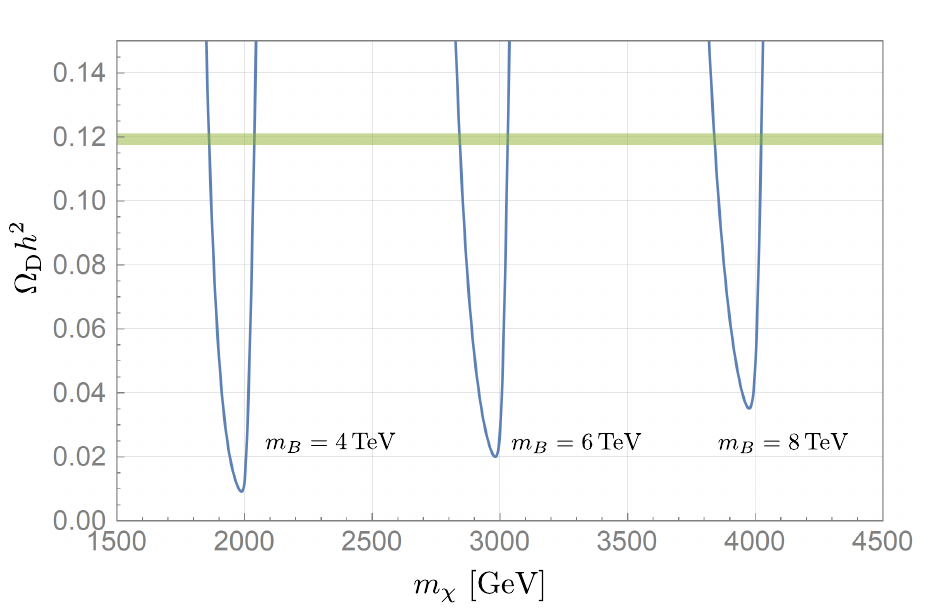}
\caption{Relic density curves for several choices of $m_B$. The horizontal band represents the observed value $\Omega_\text{D} h^2 = 0.1193 \pm 0.0009$~\cite{Workman:2022ynf}.}
\label{fig:omega-mchi} 
\end{figure}

To analyze how the choice of these mass parameters affects the predicted relic density, we scan over the gauge boson mass $m_B$ and determine the relic density as a function of  $m_\chi$; Fig.~\ref{fig:omega-mchi} shows results for three choices of $m_B$.  
A resonance effect is apparent when  $m_B \approx 2  \,m_\chi$. At this resonance, the total cross section assumes a maximum value related to the gauge boson decay width, resulting in a minimum relic density.  The value of the minimum relic density increases 
as the mass of the the U(1)$_B$ gauge boson increases, because the gauge boson width scales with $m_B$.  The observed dark matter relic density is~\cite{Workman:2022ynf} 
\begin{align} 
\Omega_\text{D} h^2 = 0.1193 \pm 0.0009 \, .
\label{eq:ohs}
\end{align} 
In Fig.~\ref{fig:omega-mchi} we superpose this band; for a given $m_B$, there are  two disconnected mass ranges for $m_\chi$ in which
a relic density consistent with observation is obtained. The allowed ranges of $m_\chi$ can be extracted as a function of $m_B$
for each asymptotically safe scenario defined in Sec.~\ref{sec:two}; these will be used in the study of the the dark matter-nucleon elastic scattering cross section 
in the next section.

\section{Direct detection} \label{sec:four}

\begin{figure}[!h]
\centering
\includegraphics[width=0.6\textwidth]{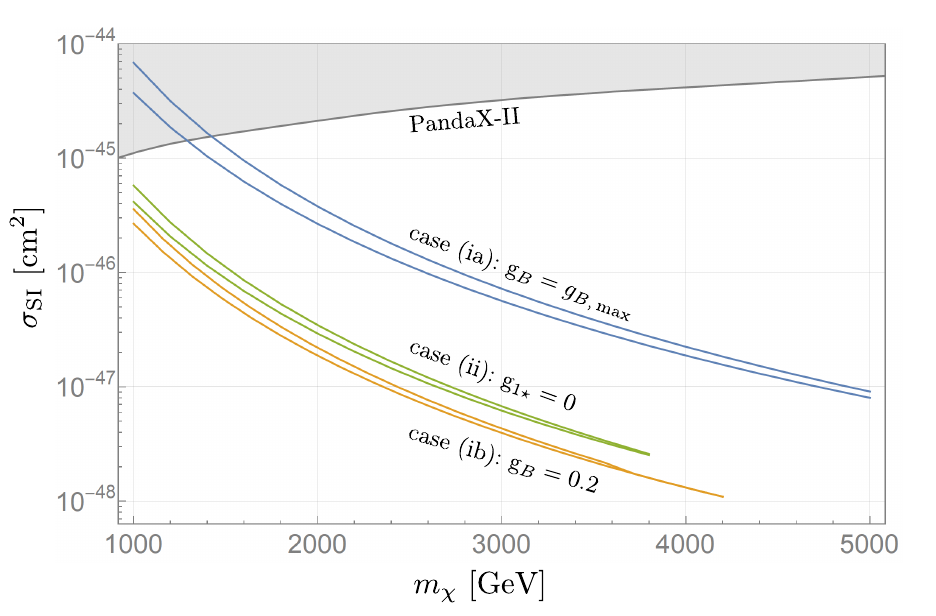}
\caption{Spin-independent dark matter-nucleon elastic scattering cross sections $\sigma_\text{SI}$, for parameter choices $(m_\chi, m_B)$ that yield the correct dark matter relic density.}
\label{fig:direct-detection} 
\end{figure}

For each point in model parameter space that leads to the correct relic density, we must check that the experimental bounds from the direct detection
of dark matter-nucleon elastic scattering are satisfied.   We consider only the most stringent bounds that follow from the spin-independent scattering
cross section.  For TeV-scale dark matter, the momentum transfer in the relevant t-channel Feynman diagrams can be neglected, $q^2 \approx 0$. The effective dimension-six 
operators, which are suppressed by $1/m _B^2$, have vector-vector, vector-axial vector, and axial vector-axial vector parts.  Only the vector-vector part, {\it i.e.} the  $\bar{\chi} \gamma^\mu \chi \bar{q} \gamma_\mu q$ operator, contributes to the spin-independent cross section~\cite{Agrawal:2010fh,Cline:2014dwa,Lin:2019}. Nucleon matrix elements of a quark vector current have form factors that simply count the
number of quarks, so there is no hadronic uncertainty in going from quark to nucleon matrix elements.  For example, elastic scattering off a nucleon $N=p$ or $n$ in our model is given by 
\begin{equation} 
\sigma_N = \frac{g_B^4}{36\, \pi} \frac{\mu_{\chi N}^2}{m_B^4} \left(1+ \sqrt{\frac{3}{5}} c_N \frac{\tilde{g}}{g_B}\right)^2 \,\,\, ,
\end{equation} 
where $c_N = 1/4$ or $3/4$ for a neutron or proton, respectively, and $\mu_{\chi N} = m_\chi m_N / (m_\chi + m_N)$ is the dark matter-nucleon reduced mass.  To compare to experimental bounds, we take into account that the dark matter scatters coherently off of the entire nucleus
so we must sum over protons and neutrons in the amplitude.  To obtain an effective dark matter-nucleon cross section we then divide the cross section by
the square of the atomic mass number.   For a Xenon target, with atomic number 54 and atomic mass 131.293, we find 
\begin{equation}
\sigma_\text{SI} = \frac{g_B^4}{36 \, \pi} \frac{\mu_{\chi N}^2}{m_B^4} \left(1+ 0.35 \, \frac{\tilde{g}}{g_B} \right)^2 \,\,\, ,
\end{equation} 
where numerically we use an average mass for the nucleon, $m_N \approx 939$ MeV.  We note that in the limit $\tilde{g} = 0$, our result agrees with the cross section given in Ref.~\cite{Duerr:2013lka};  for sample points (ia) and (ib),  $\tilde{g} = \frac{16}{77} g_B$ and the kinetic mixing
term represents a 15\% correction. In Fig.~\ref{fig:direct-detection}, for each of the asymptotically safe scenarios defined in Sec.~\ref{sec:two}, we display $\sigma_\text{SI}$ for parameter choices corresponding to relic densities within two standard deviations of the central value as per Eq.~\eqref{eq:ohs}.
We compare to the bounds from the PandaX--II experiment~\cite{Cui:2017}, which uses a Xenon detector and constrains TeV-scale dark matter masses. All three scenarios considered here are allowed by current experimental bounds for dark matter masses $m_\chi \gtrsim 1.5$~TeV.

\section{Conclusions} \label{sec:conc}
In this letter, we have modified the gauged baryon number model proposed in Ref.~\cite{Boos:2022jvc} to include a TeV-scale, fermionic dark matter candidate.   The stability of the dark matter is guaranteed by a discrete subgroup of the additional gauge symmetry, and the
new gauge boson serves as the portal between the dark and visible sectors.  The new ingredient in our study is the assumption of asymptotic safety, which reduces the space of free model parameters due to the constraint that (at least some) couplings reach nontrivial
ultraviolet fixed points.  The effect of this organizing principle is that the range of the baryon number gauge coupling at the TeV scale is constrained, and the kinetic mixing parameter at the same scale becomes a function of the baryon number gauge coupling.  This fixes the 
degree of gauge boson leptophobia once the gauge coupling of the theory is specified.   Taking into account these constraints, and including the leptonic dark matter annihilation channels that are induced by the kinetic mixing, the correct 
dark matter relic density can be obtained in a number of asymptotically safe scenarios with different patterns of ultraviolet fixed points.
For these solutions, the predicted dark matter-nucleon elastic scattering cross section is consistent with the bounds from the PandaX--II experiment~\cite{Cui:2017} which probes the dark matter masses above $1$~TeV.    Measurements
of new gauge boson properties at colliders and of the dark matter-nucleon elastic scattering cross section at direct-detection experiments may 
someday provide nontrivial tests of the relationships between couplings expected in this and other asymptotically safe gauge extensions of the 
standard model.

\begin{acknowledgments} 
We thank the NSF for support under Grants PHY-1819575 and PHY-2112460.
\end{acknowledgments}


\end{document}